\documentclass[aps,11pt]{revtex4}
\usepackage{epsfig}
\usepackage{amsmath}
\usepackage{bm}
\usepackage{times}
\usepackage{graphicx}

\def\beq{\begin{equation}}
\def\eeq{\end{equation}}
\def\bea{\begin{eqnarray}}
\def\eea{\end{eqnarray}}
\def\nn{\nonumber}
\def\Eq#1{Eq.~(\ref{#1})}

\begin{document}
\preprint{IFIC/06-30}
\title{FERMION MASSES AND THE UV CUTOFF OF THE MINIMAL REALISTIC $SU(5)$}
\author{Ilja Dor\v{s}ner}
\email{idorsner@phys.psu.ed} \affiliation{The Pennsylvania State University \\
104 Davey Lab, PMB 025, University Park, PA 16802}
\author{Pavel Fileviez P\'erez}
\email{fileviez@cftp.ist.utl.pt} \affiliation{Centro de
F{\'\i}sica Te\'orica de Part{\'\i}culas \\
Departamento de F{\'\i}sica.\ Instituto Superior T\'ecnico \\
Avenida Rovisco Pais, 1. 1049-001 Lisboa, Portugal}
\author{Germ\'an Rodrigo}
\email{german.rodrigo@ific.uv.es}
\affiliation{Instituto de F\'{\i}sica Corpuscular,
CSIC-Universitat de Val\`encia,
Apartado de Correos 22085,
E-46071 Valencia, Spain.}
\date{\today}
\begin{abstract}
We investigate the predictions for fermion masses in the minimal
realistic non-supersymmetric $SU(5)$ model with the Standard Model
matter content. The possibility to achieve $b-\tau$ unification
is studied taking into account all relevant effects. In addition,
we show how to establish an upper bound on the ultraviolet
cutoff $\Lambda$ of the theory which is compatible with the
Yukawa couplings at the grand unified scale and proton decay.
We find $\Lambda \simeq 10^{17}$\,GeV, to be considered
a conservative upper bound on the cutoff. We also
provide up-to-date values of all the fermions masses
at the electroweak scale.
\end{abstract}
\maketitle
\section{Introduction}
The hierarchy problem, unification of fundamental interactions, and
fermion mass puzzle are some of the main motivations for
physics beyond the Standard Model (SM). In particular, if we
believe in unification of electroweak and strong interactions then
the so-called grand unified theories (GUTs) represent the most
natural extensions of the Standard Model.

The first grand unified theory---the Georgi-Glashow $SU(5)$
model~\cite{GG}---was introduced in 1974. In that model each
generation of the SM matter is unified in the $\overline{\bm{5}}$
and $\bm{10}$ dimensional representations, and the minimal Higgs
sector is composed of two representations: $\bm{5}_H$ and
$\bm{24}_H$. The Georgi-Glashow model is arguably the simplest
GUT. It is very predictive but it is certainly not realistic.
Namely, one cannot unify the SM gauge couplings at the high scale,
the neutrinos are massless, and a high-scale unification of Yukawa
couplings of the down quarks and charged leptons contradicts
experimental findings for the masses of those particles.

Since the simplest $SU(5)$ GUT is not realistic it requires
appropriate modifications. There is a number of ways of doing that,
but in order to preserve the predictivity of the theory
those modifications should be minimal.
With this in mind we have proposed in Ref.~\cite{Dorsner:2005fq}
the simplest possible extension of the Georgi-Glashow model
that is in agreement with experimental observations.
Phenomenological and cosmological aspects of our proposal
have been analyzed subsequently in Ref.~\cite{Dorsner:2005ii}.

In this work we study in detail the Yukawa sector
within the proposed framework, and define the upper bound on
the ultraviolet (UV) cutoff $\Lambda$ of the theory from
the constraints imposed by proton decay lifetime measurements.
We show that this UV cutoff depends on the absolute
value of the tau lepton Yukawa coupling at the unification scale
and not on the difference of the bottom quark to the tau lepton Yukawa
couplings, as expected from bottom-tau unification.
We find $\Lambda \leq 10^{17}$\,GeV.
Furthermore, as an essential ingredient in our study we perform an
up-to-date analysis of all the fermion masses at the electroweak
scale. These values are especially relevant for numerical studies
of viability of various GUT models.

The paper is organized as follows: In Section II we describe the
minimal realistic extension of the Georgi-Glashow (GG) model. In
Section III the predictions for Yukawa couplings at the
unification scale are investigated. There we also list updated
values of all the fermion masses at the electroweak scale. The UV
cutoff of the theory is defined and evaluated in Section IV.
Finally, we conclude in Section V.

\section{The minimal realistic extension of the Georgi-Glashow model}
The Higgs sector of the minimal realistic non-supersymmetric
$SU(5)$ model~\cite{Dorsner:2005fq} is composed of the fields
$\bm{24}_H = (\Sigma_8,\Sigma_3, \Sigma_{(3,2)}, \Sigma_{(\bar{3},
2)}, \Sigma_{24}) =
(\bm{8},\bm{1},0)+(\bm{1},\bm{3},0)+(\bm{3},\bm{2},-5/6) +
(\overline{\bm{3}},\bm{2},5/6)+(\bm{1},\bm{1},0)$, $ \bm{15}_H =
(\Phi_a,\Phi_b, \Phi_c)= (\bm{1},\bm{3},1)+
(\bm{3},\bm{2},1/6)+(\bm{6},\bm{1},-2/3)$, and $\bm{5}_H = (H,
T)=(\bm{1},\bm{2},1/2)+(\bm{3},\bm{1},-1/3)$, while the matter
content remains the same as in the GG model. Here we use the SM
($SU(3) \times SU(2) \times U(1)$) decomposition to set our
notation. The Lagrangian of our model includes all possible terms
invariant under the $SU(5)$ gauge symmetry, and accordingly
includes higher-dimensional operators in order to write a
consistent relation between fermion masses. (Influence of
higher-dimensional operators on gauge coupling
constants~\cite{Hill:1983xh,Shafi:1983gz} is assumed to be
negligible.) The role of the $\bm{15}_H$ dimensional Higgs is
twofold: it generates neutrino masses through a type II see-saw
mechanism~\cite{seesaw2}, and contributes to the unification of
gauge couplings.

\begin{figure*}[ht]
\begin{center}
\epsfig{file=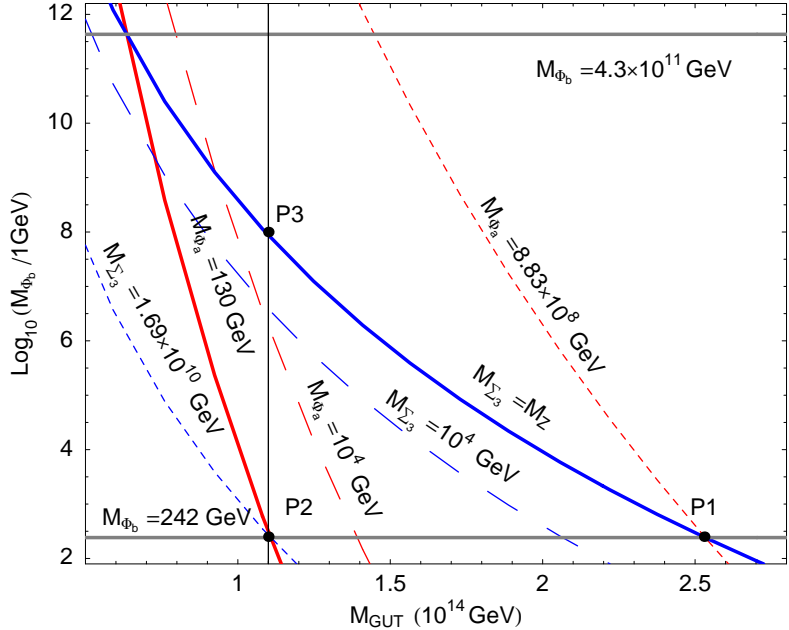,width=11cm} \caption{The whole parameter
space is shown where we can achieve gauge coupling unification.
The points P1, P2, and P3 define the allowed region, and the
corresponding boundary values for the masses of $\Phi_a$,
$\Phi_b$, and $\Sigma_3$ are shown.} \label{fig:triangle}
\end{center}
\end{figure*}

The possibility to achieve unification in the present context
has been investigated in Ref.~\cite{Dorsner:2005fq,Dorsner:2005ii}.
There we showed that there are only three fields which can help
to achieve successful unification.
Those are $\Sigma_3 \subset
\bm{24}_H$, $\Phi_a \subset \bm{15}_H$, and $\Phi_b \subset
\bm{15}_H$. We present in Fig.~\ref{fig:triangle} the appropriate
parameter space that generates successful unification of
gauge couplings at one-loop.
It corresponds to the region bounded by the lines of constant
$M_{\Phi_a}=130$\,GeV, $M_{\Phi_b}=242$\,GeV, and $M_{\Sigma_3}=M_Z$.

In fact, the maximal value of $M_{GUT}$ at the two-loop level is
somewhat larger than the one that corresponds to the benchmark
point P1 in Fig.~\ref{fig:triangle}. Namely, $M_{GUT} = 4.5 \times
10^{14}$\,GeV for $M_{\Sigma_3}=M_{\Sigma_8}=M_Z$, $M_{\Phi_a}=1.1
\times 10^4$\,GeV, $M_{\Phi_b}=242$\,GeV and
$\alpha_{GUT}^{-1}=37.1$. With this set of values we can establish
an accurate upper bound on the proton decay lifetime. In a model
independent way the proton lifetime $\tau_p$ is bounded by the
inequality~\cite{upper}:
\begin{equation}
\label{protonlifetime} \tau_p \leq 6 \times 10^{39} \,
\alpha_{GUT}^{-2} \, (M_V/10^{16} \textrm{GeV})^4 \,
(0.003 \textrm{GeV}^3/\alpha)^2\,\rm{years},
\end{equation}
where $\alpha$ is the matrix element, and $M_V$ is
a common mass of gauge bosons responsible for proton decay.
For our purposes we set $\alpha=0.015$\,GeV$^3$~\cite{Aoki:2004xe}
and identify $M_V=M_{GUT}$.
(The main source of uncertainty in Eq.~\eqref{protonlifetime}
comes from the matrix element $\alpha$. For an up-to-date discussion on
$\alpha$ see~\cite{Aoki:2006ib}. For a review on
proton stability see~\cite{review}.) Using the two-loop values
of $\alpha_{GUT}$ and $M_{GUT}$ mentioned above
we find $ \tau_p^{(\textrm{two-loops})}  \leq 1.4 \times
10^{36}$\,years~\cite{Dorsner:2005ii}. Clearly, our model could be
tested or ruled out at the next generation of proton decay
experiments~\cite{Rubbia}.

\begin{figure*}[t]
\begin{center}
\epsfig{file=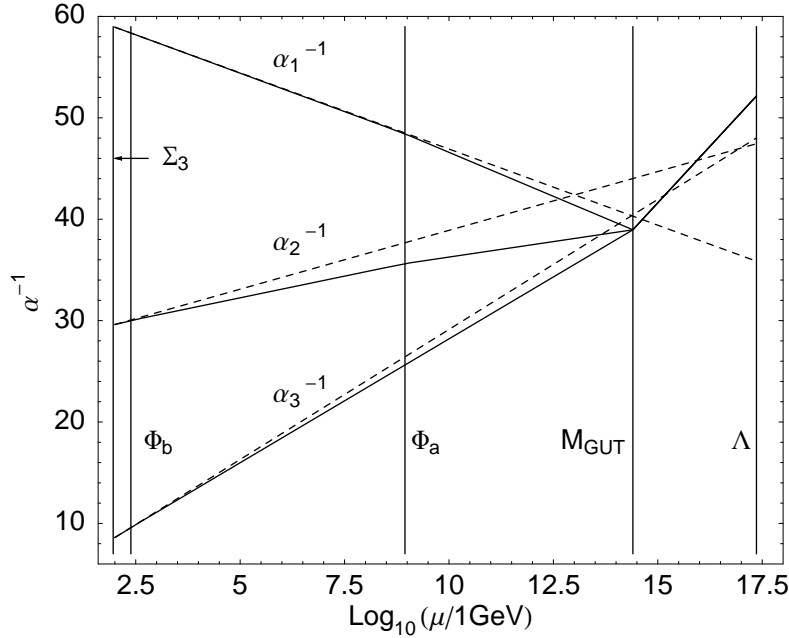,width=11cm} \caption{One-loop gauge
coupling unification at benchmark point P1 in the Minimal
Realistic SU(5) model (solid) in comparison with the SM (dashed).
Notice the asymptotic free behavior above the GUT scale.}
\label{fig:gauge}
\end{center}
\end{figure*}

Conversely, one can invert Eq.~\eqref{protonlifetime} to establish a lower
limit on the GUT scale; recall, $M_{GUT}=M_V$.
If we take $\tau_p(p \rightarrow \pi^0 e^+)>5.0 \times
10^{33}$\,years~\cite{PDG} as experimental input, and
assume $\alpha_{GUT}^{-1}\simeq 37$ the region below
$M_{GUT}=1.1 \times 10^{14}$\,GeV is excluded by proton decay.
This limit defines our benchmark point P3 in Fig.~\ref{fig:triangle}.
Once we impose proton decay constraints on $M_{GUT}$ we also get
an upper bound on the scalar leptoquark
mass from the benchmark point P3: $M_{\Phi_b} < 10^8$\,GeV.
In addition, if we consider the most natural implementation
of the type II see-saw mechanism (large $M_{\Phi_a}$) the mass of
the scalar leptoquark $\Phi_b$ comes out in the
phenomenologically interesting region ${\cal
O}(10^2-10^3)$~GeV.

We have analyzed the unification scenario at three benchmark
points P1 through P3 showed in Fig.~\ref{fig:triangle}. As we
explained, these three points define the limits of the parameter
space that yields unification of gauge couplings in the minimal
realistic $SU(5)$ model at one-loop. The particle content of the
model also implies that the gauge coupling exhibits asymptotically
free behavior between the GUT scale and the cutoff $\Lambda$. This
feature is shown in Fig.~\ref{fig:gauge} for the benchmark point
P1. The value of $\alpha_{GUT}$ at the scale $M$ ($\Lambda > M >
M_{GUT}$) is given by: $\alpha_{GUT}^{-1}(M)=
\alpha_{GUT}^{-1}(M_{GUT}) + \frac{73}{12 \pi} \ \ln
\frac{M}{M_{GUT}}$. In order to make any specific statements about
the behavior of the gauge coupling at and above the cutoff the
full structure of the more fundamental theory that represents the
ultraviolet completion of our model needs to be specified. As we
comment towards the end, there exists well-defined underlying
theory that reproduces our model below the cutoff $\Lambda$ which
also supports the asymptotic behavior of the $SU(5)$ gauge
coupling up to the Planck scale.

We have also shown that the minimal realistic $SU(5)$
predicts the existence of light fields. In the benchmark point
P1 the Higgs field $\Sigma_3$ and the scalar leptoquark $\Phi_b$ are
very light, while in the benchmark point P2 the lightest fields
are $\Phi_a$ and $\Phi_b$. Finally in the benchmark
point P3 there are two light Higgses $\Phi_a$ and $\Sigma_3$.
We can conclude then that the minimal realistic
non-supersymmetric $SU(5)$ model could be potentially tested at the
next generation of collider experiments, particularly at the Large
Hadron Collider (LHC) at CERN. The possibility to explain the
baryon-asymmetry in the Universe in this context has also been
studied~\cite{Dorsner:2005ii}. See also Ref.~\cite{ice} for the
possibility of probing second and third generation leptoquark
parameter space with the IceCube neutrino detection facility.
Since in principle all those fields could modify the predictions
for the Yukawa couplings at the unification scale we will study
their effect in the next section.

\section{Fermion masses and bottom-tau unification}

We shall investigate predictions for fermion masses in the
framework of the minimal realistic $SU(5)$ model. The RGEs for the
Yukawa couplings are summarized in Appendix B. In order to compute
the values of the Yukawa couplings at the GUT scale we need
accurate values for the fermion masses at the electroweak scale as
initial conditions for the RGEs. As a first approximation it is
sufficient to consider the third generation only, and neglect the
Yukawa couplings of the first and second generations in the RGE
evolution. For future applications however we shall update the
values of the fermion masses at the $M_Z$ scale for all the three
generations.

As input value for the top quark mass we take the latest world average
from the Tevatron Electroweak Working Group~\cite{Group:2006hz},
and extract the top quark running mass assuming that this
value corresponds to the pole mass:
\beq
\frac{M_t}{m_t(M_t)} = 1 + \frac{4}{3} \frac{\alpha_s(M_t)}{\pi} +
10.95 \left(\frac{\alpha_s(M_t)}{\pi}\right)^2 + {\cal O}(\alpha_s^3)~.
\eeq
For the bottom quark mass we will adopt a conservative
value at low energies $m_b(m_b)=4.20\pm 0.10$~GeV. This value is
compatible through QCD evolution with the experimental
measurement at higher energies~\cite{Abdallah:2005cv},
and agrees with most of the low energy determinations~\cite{PDG}.
Then, to obtain the initial conditions for \Eq{rgeyukawa}
the top and the bottom quark masses are evolved
to the $M_Z$ scale at three-loops~\cite{Rodrigo:1996gw,Rodrigo:1997zd}
with $\alpha_s(M_Z) = 0.1176 \pm 0.0020$~\cite{PDG}.
We also extract the values of the other quarks
from the PDG~\cite{PDG} and calculate them at $M_Z$
using three-loop RGE evolution in QCD accounting for
the matching conditions over the heavy quark
thresholds~\cite{Rodrigo:1996gw,Rodrigo:1997zd}.

The physical lepton masses are taken also from the PDG~\cite{PDG}.
The corresponding running masses are calculated at the $M_Z$ scale
though the relation: \beq m_{l}(M_Z) = M_{l} \left[ 1 -
\frac{\alpha(M_Z)}{\pi} \left( 1 + \frac{3}{4} \ln
\frac{M_Z^2}{M_{l}^2}\right) \right] + {\cal O}(\alpha^2)~, \eeq
with $\alpha(M_Z)^{-1}=127.906 \pm 0.019$. The input values for
all the fermion masses and the corresponding values at the
electroweak scale are summarized in Table~\ref{tab:input}.
These values at $M_Z$ substantially defer and represent
better reflection of our current knowledge of fermion masses from the
values first evaluated in Ref.~\cite{Fusaoka:1998vc}
and later updated in Ref.~\cite{Das:2000uk}.

\begin{table}[th]
\caption{\label{tab:input}
Input parameters for the fermion masses and
their values at the $M_Z$ scale. Capital $M$ denotes pole masses,
while $m(\mu)$ are running masses.}
\begin{tabular}{ccc}
\hline
$\qquad \qquad$ & $\qquad \qquad \qquad$ input value $\qquad \qquad \qquad$
& $\qquad \qquad$ running mass at $M_Z$ $\qquad \qquad$ \\ \hline
$t$ & $M_t= 171.4 \pm 2.1$ GeV       & $m_t(M_Z) = 170.3 \pm 2.4$  GeV\\
$b$ & $m_b(m_b) = 4.20 \pm 0.10$ GeV & $m_b(M_Z) =  2.89 \pm 0.11$ GeV\\
$c$ & $m_c(m_c) = 1.25 \pm 0.09$ GeV & $m_c(M_Z) =  0.63 \pm 0.08$ GeV\\
$s$ & $m_s(2\mathrm{GeV}) =  95 \pm 25$ MeV  & $m_s(M_Z) =   56 \pm 16 $ MeV\\
$u$ & $m_u(2\mathrm{GeV}) =  2.3 \pm 0.8$ MeV & $m_u(M_Z) =  1.4 \pm 0.5$ MeV\\
$d$ & $m_d(2\mathrm{GeV}) =  5.0 \pm 2.0$ MeV & $m_d(M_Z) =  3.0 \pm 1.2$ MeV\\
$\tau$ &$M_{\tau} = 1776.99^{+0.29}_{-0.26}$ MeV & $m_{\tau}(M_Z) =
1746.45^{+0.29}_{-0.26}$ MeV  \\
$\mu$ & $M_{\mu}  = 105.6583692(94)$ MeV   & $m_{\mu}(M_Z) = 102.72899(44)$ MeV  \\
$e$ & $M_{e}    =   0.510998918(44)$ MeV & $m_{e}(M_Z)   = 0.4866613(36)$ MeV  \\
\hline
\end{tabular}
\end{table}


Let us now study the predictions for fermion masses at the GUT scale.
The leptoquark $\Phi_b$ contributes at one-loop to the running
of the Yukawa couplings for charged leptons and down quarks, while the field
$\Phi_a$ modifies the RGEs for charged leptons.
The field $\Sigma_3$ does not couple to matter, only to the SM Higgs.
It contributes to the renormalization of the mass and the
couplings of the SM Higgs but does not modify the RGEs of the
Yukawa couplings. The equations for the running of the Yukawa
matrices $Y_E$ and $Y_D$ are given by:
\bea
&& 16 \pi^2 \, \frac{d Y_D}{d \ln \mu } =
Y_D \beta^{\rm{SM}}_{D} + Y_{\nu} Y_{\nu}^{\dagger} Y_D
\ \Theta(\mu - M_{\Phi_b})~,
\nn \\
&& 16 \pi^2 \, \frac{d Y_E}{d \ln \mu}= Y_E \left(\beta^{\rm{SM}}_{E}
+ \frac{3}{4} Y_{\nu}^{\dagger}  Y_{\nu} \ \Theta(\mu - M_{\Phi_a})
+ \frac{3}{2} Y_{\nu}^{\dagger} Y_{\nu} \ \Theta(\mu - M_{\Phi_b})
\right)~,
\label{rgeyukawasu5}
\eea
where $\beta^\mathrm{SM}_i$ are the SM beta coefficients (see \Eq{betayukawa1}).
The RGE of the up quark Yukawas is not modified with respect to the SM.
The contributions of the fields $\Phi_a$ and $\Phi_b$ is due to
the operator $Y_{\nu} \bar 5 \bar 5 \ 15_H$ of the Yukawa potential
that generates the interactions $Y_{\nu} \bar{l}_L^C \Phi_a \l_L$ and
$Y_{\nu} \bar{d}_R \Phi_b l_L$. In the above equation the contribution
of the field $\Phi_a$ has been taken from Ref.~\cite{RGEs-typeII}.
Notice that usually in $SU(5)$ theories there is no relationship
between the Yukawa couplings of neutrinos and charged fermions.

\begin{figure*}[th]
\begin{center}
\epsfig{file=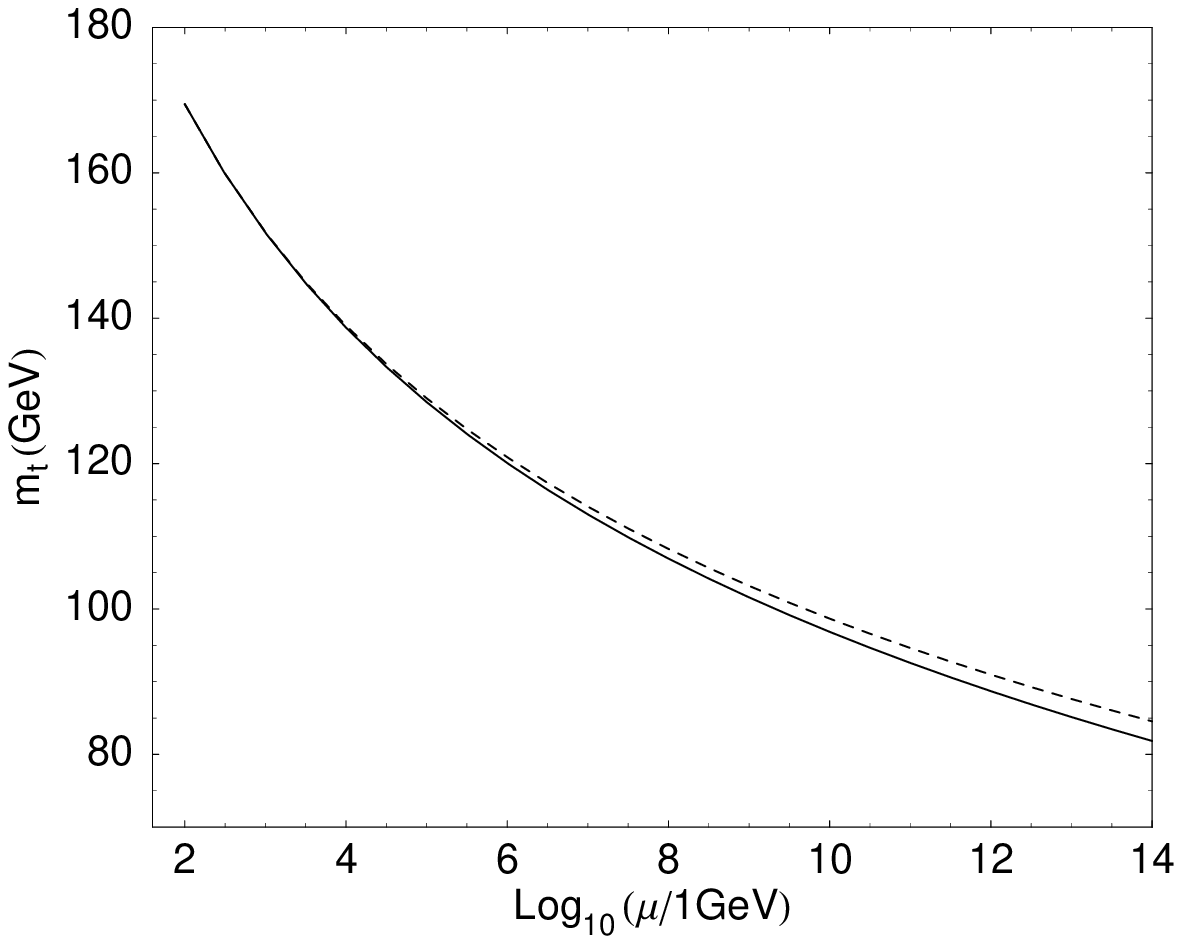,width=8cm}
\epsfig{file=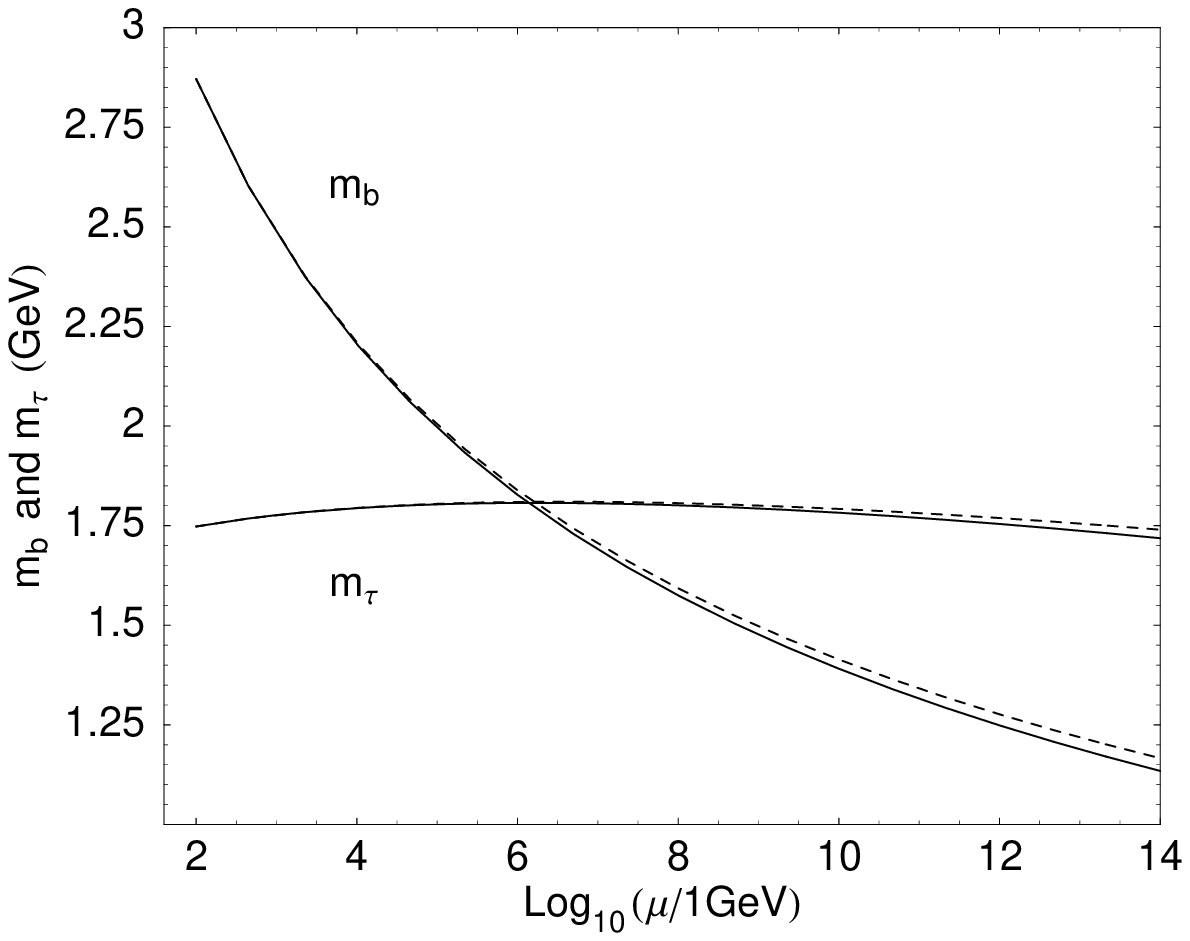,width=8cm} \caption{One-loop
evolution of the masses of the third generation in the benchmark
point P1 of the minimal realistic $SU(5)$ model (solid) and in the
SM (dashed).} \label{fig:taubottom}
\end{center}
\end{figure*}

Let us analyze the scenario where the contributions of
the field $\Phi_a$ and the leptoquark $\Phi_b$ have been
neglected in~\Eq{rgeyukawasu5}.
Notice that at one-loop the mass of the field $\Phi_a$ in all
the parameter space is much below the natural value for the
see-saw mechanism ($M_{\Phi_a}=10^{13}-10^{14}$\,GeV),
and the Yukawa couplings for neutrinos are expected to be small.
In Fig.~\ref{fig:taubottom}
we show the one-loop evolution of the top, bottom and tau
masses respectively in comparison with the SM for the benchmark point P1.
Since in this case we have neglected the neutrino Yukawa contributions,
the RGEs in the minimal realistic $SU(5)$ model are the same as in the SM
(\Eq{rgeyukawa}), and the evolution of the charged fermion masses
in that model is modified with respect to the SM only through the
change in the evolution of the gauge couplings.
Consequently, we obtain values of
the bottom and tau Yukawa couplings at the GUT scale that
are only 1--2\% away from the SM prediction. A similar situation
will happen for the other two benchmark points.
Thus, independently of the benchmark point the Yukawa coupling
of the bottom quark will lie at high energies below the Yukawa
coupling of the tau lepton, and therefore it is not possible to achieve
unification of $Y_\tau$ and $Y_b$.

If we include the contributions of $\Phi_a$ and/or
the leptoquark $\Phi_b$ in the running the difference
between $Y_\tau$ and $Y_b$ at the GUT scale will be increased.
The reason is that both contributions are positive, and
the coefficient of the neutrino Yukawa coupling
in the RGE for $Y_E$ is larger than that for $Y_D$.
This will result into a larger tau Yukawa coupling, and as
we will see in the next Section into an smaller UV cutoff.
In order to obtain a conservative upper bound on the UV cutoff
we will focus our analysis on the first scenario where the extra
contributions of the fields $\Phi_a$ and $\Phi_b$ are neglected.

\section{The UV cutoff of the theory and proton decay}

In this section we study the possibility to establish an upper
bound on the UV cutoff of the minimal realistic $SU(5)$ model.
The relevant Yukawa potential up to
order $1/\Lambda$ is defined by~\cite{Dorsner:2005fq}:
\begin{eqnarray}
V_\mathrm{Yukawa}&=& \epsilon_{ijklm} \left( 10_a^{ij} \ Y_{ab} \
10_b^{kl} \ 5_H^m \ + \ 10_a^{ij} \ Y^{(1)}_{ab} \ 10^{kl}_b \
\frac{(24_H)^m_n}{\Lambda} 5_H^n \
+ \ 10_a^{ij} \ Y^{(2)}_{ab} \ 10^{kn}_b \ 5_H^l \ \frac{(24_H)^m_n}{\Lambda}\right) \nonumber \\
&+& {5_H^*}_i \ 10^{ij}_a \ Y^{(3)}_{ab} \ \bar{5}_{bj} \ + \
{5_H^*}_i \ \frac{(24_H)^i_j}{\Lambda} \ 10^{jk}_a \ Y^{(4)}_{ab}
\ \bar{5}_{bk}
\ + \ {5_H^*}_i \ 10^{ij}_a \ Y^{(5)}_{ab} \ \frac{(24_H)^k_j}{\Lambda} \ \bar{5}_{bk} \nonumber \\
\label{potential} &+& \ \bar{5}_{ai} \ Y^{(6)}_{ab} \ \bar{5}_{bj}
\ 15_H^{ij} \ + \ \bar{5}_{ai} \ Y^{(7)}_{ab} \ \bar{5}_{bj} \
\frac{5_H^i \ 5_H^j}{\Lambda}~,
\end{eqnarray}
where $i,j,k,l,m$ and $n$ are the $SU(5)$ indices, while $a,b$ and
$c$ are the family indices. Once $\bm{24}_H$ gets a vacuum
expectation value (VEV) $\langle \bm{24}_H \rangle= \sigma \,
\textrm{diag}(2,2,2,-3, -3)$ the GUT symmetry $SU(5)$ is broken to
the SM gauge symmetry. Then, the Yukawa couplings for charged
fermions read~\cite{Bajc}:
\begin{eqnarray}
Y_U &=& 4 (Y + Y^T)  - 12 \frac{\sigma}{\Lambda} ( Y^{(1)} +
{Y^{(1)}}^T) -  2 \frac{\sigma}{\Lambda} ( 4 Y^{(2)} -
{Y^{(2)}}^T )~,
\\ \label{YD}
Y_D &=& - Y^{(3)} \ + \frac{\sigma}{\Lambda} ( 3 Y^{(4)} - 2
Y^{(5)})~,
\\ \label{YE}
Y_E &=& - Y^{(3)} \ + 3 \frac{\sigma}{\Lambda} ( Y^{(4)} +
Y^{(5)})~.
\end{eqnarray}
For neutrino masses, on the other hand, we find:
\begin{equation}
M_{\nu} = Y^{(6)} \langle \delta^0 \rangle
+  Y^{(7)} \frac{\langle H^0 \rangle^2}{\Lambda}~,
\end{equation}
where $\langle \delta^0\rangle$ and $\langle H^0 \rangle$ are the
VEVs of the neutral components of $\Phi_a$ and $H$, respectively.
The Yukawa couplings for charged fermions are diagonalized as
follows: $U^T \ Y_U \ U_C = Y_U^\mathrm{diag}$, $D^T \ Y_D \ D_C =
Y_D^\mathrm{diag}$ and $E^T_C \ Y_E \ E = Y_E^\mathrm{diag}$.

Using the relevant relations for the Yukawa couplings we find:
\begin{equation}
Y_E - Y_D = Y^{(5)} \frac{1}{\sqrt{ 2 \pi
\alpha_{GUT}}}\frac{M_V}{\Lambda},
\end{equation}
where we use $\sigma=\lambda/\sqrt{30}$ and
$M_V=M_{GUT}=\sqrt{\frac{5}{12}} \ \lambda \ g_{GUT}$ with
$g_{GUT}^2= 4 \pi \alpha_{GUT}$. For the physical Yukawa couplings
the relation reads:
\begin{equation}
\label{Yed} Y_E^\mathrm{diag} = E_C^T D^* Y_D^\mathrm{diag} D_C^{\dagger} E \ +
\ E_C^T Y^{(5)} E \frac{M_V}{{\sqrt{2\pi \alpha_{GUT} }}\Lambda}.
\end{equation}
If we require that the theory remains perturbative at the GUT
scale: $|c_{kk}|=|(E_C^T Y^{(5)} E)_{kk}|/\sqrt{4 \pi}
\leq 1$; hence the upper bound on
$\Lambda$ can be parametrized as:
\begin{equation}
\label{Upper} \Lambda \ \leq \ \sqrt{\frac{2}{\alpha_{GUT}}}
\times \frac{M_{GUT}} {|Y_{\tau} - a^{3i} Y_D^{i}|}~.
\end{equation}
Here $a^{ki}=(E_C^T D^*)^{ki}(D_C^{\dagger} E)^{ik}$.
We have set $k=3$ because in that case one finds the strongest
upper bound. Notice that when $a^{3i} > 0$ ($a^{3i} <
0$) and real, the maximal value of the denominator in
Eq.~(\ref{Upper}) is $Y_{\tau} - Y_d$ ($Y_{\tau} + Y_b $).
Since the GUT scale in our model is very low we have to be
sure that it is possible to satisfy the experimental constraints
on the proton decay lifetime. In particular, before defining the
upper bound on the UV cutoff we have to determine which is the
pattern of the $a^{ki}$ coefficients that is in agreement with
the constraints set by nucleon decay.

Let us discuss the relation between the upper bound on $\Lambda$
and proton decay. It was pointed out in Ref.~\cite{upper} that
in order to find the upper bound on the proton decay lifetime in
the context of the minimal realistic $SU(5)$ model the following conditions
have to be fulfilled:
\begin{equation}
E_C = D \ B_1~, \qquad D_C = E \ B_2~,
\end{equation}
where
\begin{equation}
B_{j}=\left(
\begin{array}{ccc}
  0 & 0 & e^{i\alpha_j} \\
  0 & e^{i\beta_j} & 0 \\
  e^{i\gamma_j} & 0 & 0 \\
\end{array}\right)~.
\end{equation}
In this case we have the maximal suppression for the proton decay
channels. Now, using the above
relations we get $c_{ii}= (B_1^T D^T Y^{(5)} E)_{ii} /
\sqrt{4\pi}$ and the upper bound on $\Lambda$ (neglecting the
phases) is given by:
\begin{equation}
\label{upper-real} \Lambda \ \leq \ \sqrt{\frac{2}{\alpha_{GUT}}}
\times \frac{M_{GUT}}{|Y_{\tau}- Y_d|}~.
\end{equation}
Notice that this is the bound which corresponds to the case
$a^{3i} > 0$. By definition it is consistent with fermion masses
and proton decay simultaneously. Therefore, using the values of the Yukawa
couplings at the GUT scale, the most conservative upper bound on
$\Lambda$ is coming from the scenario when we have maximal suppression
of the proton decay channels. It is important to say that usually
the cutoff of a GUT model is evaluated from the difference
between $Y_b$ and $Y_\tau$~\cite{Berezhiani}. However, here we have
shown that the most conservative upper bound on the UV cutoff,
as given by Eq.~(\ref{upper-real}), is defined by the difference
between $Y_\tau$ and $Y_d$, and not by the departure from
$b-\tau$ unification.

In the previous section we have already addressed the issue of
numerical values of Yukawa couplings at the GUT scale. Since the
value of the tau Yukawa coupling is almost independent of the benchmark
point, the actual bound on the cutoff simply depends on the ratio
$M_{GUT}/\sqrt{\alpha_{GUT}}$. We hence summarize
in Table~\ref{tab:bench} the values of the masses of the fields
$\Phi_a$, $\Phi_b$, and $\Sigma_3$ at each benchmark
point and the corresponding values of $\alpha_{GUT}$, $M_{GUT}$,
$Y_\tau(M_{GUT})$, and the ultraviolet cutoff $\Lambda$.
Notice that the most conservative upper
bound on the cutoff of the theory at one-loop level is
$\Lambda^{upper}\simeq10^{17}$\,GeV, which could be identified
with the string unification scale~\cite{Dienes}.

\begin{table}[th]
\caption{\label{tab:bench} UV Cutoff for the three benchmark
points. All the mass scales in GeV}
\begin{tabular}{cccccccc} \hline
benchmark point & $M_{\Phi_a}$ & $M_{\Phi_b}$ & $M_{\Sigma_3}$ &
$\alpha_{GUT}^{-1}$ & $M_{GUT}$ & $Y_\tau(M_{GUT})$ & $\Lambda$ \\
\hline
P1 & $8.83 \times 10^8$ & $242$ & $M_Z$ & 38.9 & $2.53\times 10^{14}$ & $0.0098$ & $2.3 \times 10^{17}$ \\
P2 & $130$ & $242$     & $1.69 \times 10^{10}$ & 38.1& $1.11\times 10^{14}$ & $0.0098$ & $1.0 \times 10^{17}$ \\
P3 & $6.92 \times 10^4$ & $8.68 \times 10^7$ & $M_Z$ & 38.7&
$1.10\times 10^{14}$ & $0.0099$ & $9.8 \times 10^{16}$ \\ \hline
\end{tabular}
\end{table}

We briefly investigate how our analysis holds at the two-loop
level. In Fig.~\ref{fig:triangle}, we show our results for the
scenario which corresponds to the benchmark point P1.
To insure the proper inclusion of boundary
conditions~\cite{Hall:1980kf} at $M_{GUT}$ we set
$\left.\alpha^{-1}_i\right|_{GUT}=\alpha^{-1}_{GUT}-\lambda_i/(12
\pi)$, where $\{\lambda_1,\lambda_2,\lambda_3\}=\{5,3,2\}$. (In
addition to the central values given in Table~\ref{tab:input} we
use CKM angles and phases from~\cite{PDG} and $\alpha_3(M_Z) =
0.1176 \pm 0.0020$, $\alpha_2(M_Z) = 0.033816 \pm 0.000027$ and
$\alpha_1(M_Z) = 0.016949 \pm 0.000005$ as our input.)

\begin{figure}[th]
\begin{center}
\includegraphics[width=5in]{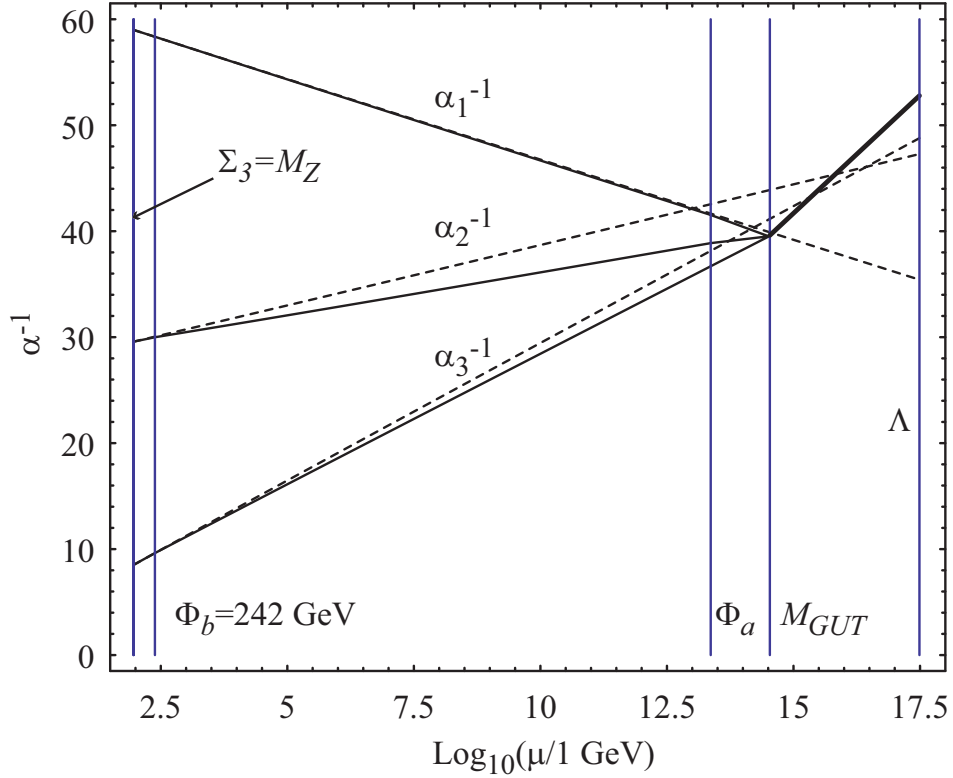}
\end{center}
\caption{\label{figure:CutOffReal} The gauge coupling unification
at the two-loop level for central values of low-energy
observables. The two-loop SM running is presented by dashed lines.
Solid lines correspond to the benchmark scenario P1 with
$\Sigma_3$, $\Phi_b$ and $\Phi_a$ below the GUT scale. Vertical
lines mark the relevant scales: $M_{\Sigma_3}=M_Z$,
$M_{\Phi_b}=242$\,GeV, $M_{\Phi_a}=2.3 \times 10^{13}$\,GeV,
$M_{GUT}=3.4 \times 10^{14}$\,GeV and $\Lambda=3.1 \times
10^{17}$\,GeV.}
\end{figure}

Comparison between Figs.~\ref{fig:gauge}
and~\ref{figure:CutOffReal} shows that the GUT scale is slightly
larger in the two-loop case. On the other hand, $\Phi_a$ is
significantly larger than in the one-loop case. And, we get
$Y_{\tau}=0.00986$ and accordingly $\Lambda=3.1 \times
10^{17}$\,GeV for the benchmark point P1, slightly above
the one-loop result.


Before we summarize our results, let us address the issue of the
possible origin of higher-dimensional operators we invoke in
Eq.~\eqref{potential}. After all, they play decisive role in
making our $SU(5)$ model realistic. Namely, they affect the Yukawa
sector in a way that simultaneously allows for realistic charged
fermion masses and efficient suppression of the gauge mediated
proton decay. In particular, they modify the GUT scale relation
between $Y_E$ and $Y_D$ that basically rules out the GG model.
And, they violate another prediction of the GG model, i.e.,
$Y_U=Y^T_U$, which would prevent one from completely suppressing
the gauge boson mediated proton decay~\cite{upper}. In fact, our
considerations of these modifications and proton decay constraints
resulted in the upper bound on the cutoff $\Lambda$.

As we have shown, the cutoff comes out to be significantly below
the Planck scale. This means that we cannot resort to the Planck
scale effects to generate necessary higher-dimensional operators.
It is then natural to ask for the credible renormalizable model
that would effectively mimic the original proposal below the
cutoff. To this end we observe that the most minimal
renormalizable setup that yields the original model requires
introduction of the following two matter pairs:
($\bm{5}$,$\overline{\bm{5}}$) and
($\bm{10}$,$\overline{\bm{10}}$). These pairs can clearly have
gauge invariant mass terms above the GUT scale that can be
identified with the scale $\Lambda$. Once these fields are
integrated out the effective model below $\Lambda$ would have
exactly the same features as the original model considered in this
paper. For example, the relevant operators that eventually modify
$Y_E=Y_D$ relation are $a_i \bm{10}_i \bm{5}^\dagger_H
\overline{\bm{5}}$ and $b_i \overline{\bm{5}}_i \bm{24}_H \bm{5}$.
($\bm{10}$,$\overline{\bm{10}}$) pair is needed to modify
$Y_U=Y^T_U$. Clearly, such a simple renormalizable realization of
the model would also imply asymptotic freedom of the $SU(5)$ gauge
coupling between $\Lambda$ and the Planck scale.

\section{Summary}
We have investigated the predictions for fermion masses in the
minimal realistic non-supersymmetric grand unified model with
the SM matter content based on $SU(5)$ gauge symmetry.
We have shown that it is not possible to
achieve $b-\tau$ unification in this context since the extra
contributions to the running of the Yukawa couplings are
always positive and larger for $Y_E$. We pointed out
that the upper bound on the ultraviolet cutoff of the
theory is $\Lambda^{upper} \simeq 10^{17}$\,GeV which is
consistent with the predictions for Yukawa
couplings and the constraints coming from proton decay.
In addition, we have provided up-to-date values of
all the fermions masses at the electroweak scale.
\section*{Acknowledgements}

{\small The work of G.~R was partially supported by Ministerio de
Educaci\'on y Ciencia (MEC) under grant FPA2004-00996, and
Generalitat Valenciana under grant GV05-015, and
Consejo Superior de Investigaciones Cient\'{\i}ficas (CSIC)
under grant PIE 200650I247. P.~F.~P has been
supported by {\em Funda\c{c}\~{a}o para a Ci\^{e}ncia e a
Tecnologia} (FCT, Portugal) through the project CFTP,
POCTI-SFA-2-777 and a fellowship under project
POCTI/FNU/44409/2002. P.~F.~P would like to thank the Instituto de
F\'{\i}sica Corpuscular (IFIC) in Valencia for hospitality.}

\appendix
\section{RGEs of the gauge couplings at two-loops}
The relevant equations for the running of the gauge couplings at
the two-loop level take the form
\begin{equation}
\frac{\textrm{d}\,\alpha_i(\mu)}{\textrm{d}\, \ln \mu}
=\frac{b_i}{2 \pi}\, \alpha^2_i(\mu)+\frac{1}{8 \pi^2}
\sum_{j=1}^{3} b_{ij}\, \alpha^2_i(\mu)\,\alpha_j(\mu)+\frac{1}{32
\pi^3}\, \alpha^2_i(\mu) \sum_{l=U,D,E} \text{Tr}\,[C_{il}
Y^\dagger_l Y_l]\,.
\end{equation}
The general formula for $b_i$ and $b_{ij}$ coefficients is given
in~\cite{Jones:1981we}. Besides the well-known SM coefficients we
have:
\begin{equation*}
b_{i}^{\Sigma_3}= \left(\begin{array}{c}
   0 \\
   \frac{1}{3} \\
   0 \\
\end{array}\right),
\qquad b_{i}^{\Sigma_8}= \left(\begin{array}{c}
   0 \\
   0 \\
   \frac{1}{2} \\
\end{array}\right),
\qquad b_{i}^{\Phi_b}= \left(\begin{array}{c}
   \frac{1}{30} \\
   \frac{1}{2} \\
   \frac{1}{3} \\
\end{array}\right),
\qquad b_{i}^{\Phi_a}= \left(\begin{array}{c}
   \frac{3}{5} \\
   \frac{2}{3} \\
   0 \\
\end{array}\right),
\end{equation*}
\begin{equation*}
b_{ij}^{\Sigma_3}=\left(
\begin{array}{ccc}
  0 & 0 & 0 \\
  0 & \frac{28}{3} & 0 \\
  0 & 0 & 0 \\
\end{array}\right),
\quad b_{ij}^{\Sigma_8}=\left(
\begin{array}{ccc}
  0 & 0 & 0 \\
  0 & 0 & 0 \\
  0 & 0 & 21 \\
\end{array}\right),
\quad b_{ij}^{\Phi_b}=\left(
\begin{array}{ccc}
  \frac{1}{150} & \frac{3}{10} & \frac{8}{15} \\
  \frac{1}{10} & \frac{13}{2} & 8 \\
  \frac{1}{15} & 3 & \frac{22}{3} \\
\end{array}\right),
\quad b_{ij}^{\Phi_a}=\left(
\begin{array}{ccc}
  \frac{108}{25} & \frac{72}{5} & 0 \\
  \frac{24}{5} & \frac{56}{3} & 0 \\
  0 & 0 & 0 \\
\end{array}\right),
\end{equation*}
which we incorporate at the appropriate scales. The $C_{il}$
coefficients are~\cite{Arason:1991ic}:
\begin{equation*}
C_{il}=\left(
\begin{array}{ccc}
  \frac{17}{10} & \frac{1}{2} & \frac{3}{2} \\
  \frac{3}{2} & \frac{3}{2} & \frac{1}{2} \\
  2 & 2 & 0 \\
\end{array}\right).
\end{equation*}
Obviously, Yukawa couplings enter the gauge coupling running at
the two-loop level. Thus, one needs to run them as well at the
one-loop level for consistency. We use the SM one-loop equations
for the Yukawa couplings that can be found, for example, in
Ref.~\cite{Arason:1991ic}.
\section{RGEs of the Yukawa couplings}
The beta functions for the evolution of the Yukawa couplings in the SM are given by
\bea
\beta_U^\mathrm{SM} &=& T-G_U+\frac{3}{2} (Y^\dagger_U Y_U -Y^\dagger_D Y_D )~, \nn \\
\beta_D^\mathrm{SM} &=& T-G_D+\frac{3}{2} (Y^\dagger_D Y_D -Y^\dagger_U Y_U )~, \nn \\
\beta_E^\mathrm{SM} &=& T-G_E+\frac{3}{2} \, Y^\dagger_E Y_E ~,
\label{betayukawa1}
\eea
where
\beq
T=\mathrm{Tr} \left( Y^\dagger_E Y_E + 3 Y^\dagger_U Y_U +3 Y^\dagger_D Y_D \right)~,
\eeq
and
\beq
\left(
\begin{array}{c} G_U \\ G_D \\ G_E
\end{array}\right) = \left(
\begin{array}{ccc}
  \frac{17}{20} & \frac{9}{4} & 8 \\
  \frac{1}{4}   & \frac{9}{4} & 8 \\
  \frac{9}{4}   & \frac{9}{4} & 0 \\
\end{array}\right)\left(
\begin{array}{c} g_1^2 \\ g_2^2 \\ g_3^2
\end{array}\right).
\eeq
It is convenient~\cite{Babu:1987im} to define the following matrices:
\beq
M_E = Y^\dagger_E Y_E~, \qquad M_D = Y^\dagger_D Y_D~,
\qquad M_U = Y^\dagger_U Y_U~,
\eeq
whose diagonalization is given by
\bea
E^\dagger \, M_E \, E = \mathrm{diag} \left(Y_e^2,Y_\mu^2,Y_\tau^2 \right)~, \nn \\
D^\dagger_C \, M_D \, D_C = \mathrm{diag} \left(Y_d^2,Y_s^2,Y_b^2 \right)~, \nn \\
U^\dagger_C \, M_U \, U_C = \mathrm{diag}
\left(Y_u^2,Y_c^2,Y_t^2 \right)~, \eea The CKM matrix is defined
through $V^\mathrm{CKM}=U^\dagger_C {D_C}$. Taking into account
that $A A^{\dagger}=1$, for $A=U,D,E$, one can derive RGEs
for the diagonal elements of these matrices: \bea 4\pi \, \frac{d
\alpha_i^U}{d\ln \mu^2} &=& \alpha_i^U \bigg[ \bar T - \bar G_U +
\frac{3}{2} \alpha_i^U - \frac{3}{2} \sum_j
|V_{ij}^\mathrm{CKM}|^2  \alpha_j^D
\bigg]~, \nn \\
4\pi \, \frac{d \alpha_j^D}{d \ln \mu^2} &=& \alpha_j^D
\bigg[ \bar T - \bar G_D + \frac{3}{2} \alpha_j^D - \frac{3}{2} \sum_i |V_{ij}^\mathrm{CKM}|^2  \alpha_i^U
\bigg]~, \nn \\
4\pi \, \frac{d \alpha_i^E}{d \ln \mu^2} &=& \alpha_i^E
\bigg[ \bar T - \bar G_E + \frac{3}{2} \alpha_i^E \bigg]~,
\label{rgeyukawa}
\eea
where $\bar T=T/(4\pi)$, $\bar G_l = G_l/(4\pi)$
and $\alpha_i^l=(Y_i^l)^2/(4\pi)$, with $i=1,2,3$ the family index.


\end{document}